%




\font\bigbf=cmbx10 scaled\magstep2

\font\twelverm=cmr10 scaled 1200    \font\twelvei=cmmi10 scaled 1200
\font\twelvesy=cmsy10 scaled 1200   \font\twelveex=cmex10 scaled 1200
\font\twelvebf=cmbx10 scaled 1200   \font\twelvesl=cmsl10 scaled 1200
\font\twelvett=cmtt10 scaled 1200   \font\twelveit=cmti10 scaled 1200

\skewchar\twelvei='177   \skewchar\twelvesy='60


\def\twelvepoint{\normalbaselineskip=12.4pt
  \abovedisplayskip 12.4pt plus 3pt minus 9pt
  \belowdisplayskip 12.4pt plus 3pt minus 9pt
  \abovedisplayshortskip 0pt plus 3pt
  \belowdisplayshortskip 7.2pt plus 3pt minus 4pt
  \smallskipamount=3.6pt plus1.2pt minus1.2pt
  \medskipamount=7.2pt plus2.4pt minus2.4pt
  \bigskipamount=14.4pt plus4.8pt minus4.8pt
  \def\rm{\fam0\twelverm}          \def\it{\fam\itfam\twelveit}%
  \def\sl{\fam\slfam\twelvesl}     \def\bf{\fam\bffam\twelvebf}%
  \def\mit{\fam 1}                 \def\cal{\fam 2}%
  \def\tt{\twelvett}
  \textfont0=\twelverm   \scriptfont0=\tenrm   \scriptscriptfont0=\sevenrm
  \textfont1=\twelvei    \scriptfont1=\teni    \scriptscriptfont1=\seveni
  \textfont2=\twelvesy   \scriptfont2=\tensy   \scriptscriptfont2=\sevensy
  \textfont3=\twelveex   \scriptfont3=\twelveex 
 \scriptscriptfont3=\twelveex
  \textfont\itfam=\twelveit
  \textfont\slfam=\twelvesl
  \textfont\bffam=\twelvebf \scriptfont\bffam=\tenbf
  \scriptscriptfont\bffam=\sevenbf
  \normalbaselines\rm}



\def\beginlinemode{\endmode
  \begingroup\parskip=0pt 
\obeylines\def\\{\par}\def\endmode{\par\endgroup}}
\def\beginparmode{\endmode
  \begingroup \def\endmode{\par\endgroup}}
\let\endmode=\par
{\obeylines\gdef\
{}}
\def\singlespace{\baselineskip=\normalbaselineskip}
\def\oneandathirdspace{\baselineskip=\normalbaselineskip
  \multiply\baselineskip by 4 \divide\baselineskip by 3}
\def\oneandahalfspace{\baselineskip=\normalbaselineskip
  \multiply\baselineskip by 3 \divide\baselineskip by 2}
\def\doublespace{\baselineskip=
\normalbaselineskip \multiply\baselineskip by 2}

\newcount\firstpageno
\firstpageno=1
\footline={\ifnum\pageno<\firstpageno{\hfil}%
\else{\hfil\twelverm\folio\hfil}\fi}
\let\rawfootnote=\footnote              
\def\footnote#1#2{{\rm\singlespace\parindent=0pt\rawfootnote{#1}{#2}}}
\def\raggedcenter{\leftskip=4em plus 12em \rightskip=\leftskip
  \parindent=0pt \parfillskip=0pt \spaceskip=.3333em \xspaceskip=.5em
  \pretolerance=9999 \tolerance=9999
  \hyphenpenalty=9999 \exhyphenpenalty=9999 }
\def\dateline{\rightline{\ifcase\month\or
  January\or February\or March\or April\or May\or June\or
  July\or August\or September\or October\or November\or December\fi
  \space\number\year}}
\def\received{\vskip 3pt plus 0.2fill
 \centerline{\sl (Received\space\ifcase\month\or
  January\or February\or March\or April\or May\or June\or
  July\or August\or September\or October\or November\or December\fi
  \qquad, \number\year)}}


\hsize=6.5truein
\vsize=8.9truein
\voffset=0.0truein
\parskip=\medskipamount
\twelvepoint            
\oneandathirdspace           
\overfullrule=0pt       



\def\title                      
  {\null\vskip 3pt plus 0.2fill
   \beginlinemode \doublespace \raggedcenter \bigbf}

\def\author                     
  {\vskip 3pt plus 0.2fill \beginlinemode
   \singlespace \raggedcenter}

\def\affil                      
  {\vskip 4pt 
\beginlinemode
   \singlespace \raggedcenter \sl}

\def\abstract                   
  {\vskip 3pt plus 0.3fill \beginparmode
   \oneandathirdspace\narrower}

\def\endtitlepage               
  {\endpage                     
   \body}

\def\body                       
  {\beginparmode}               

\def\head#1{                    
  \vskip 0.25truein     
 {\immediate\write16{#1}
   \noindent{\bf{#1}}\par}
   \nobreak\vskip 0.01truein\nobreak}

\def\subhead#1{                 
  \vskip 0.25truein             
  \noindent{{\it {#1}} \par}
   \nobreak\vskip 0.005truein\nobreak}

\def\refto#1{[#1]}           

\def\references                 
  {\subhead{\bf References}         
   \beginparmode
   \frenchspacing \parindent=0pt \leftskip=1truecm
   \doublespace\parskip=8pt plus 3pt
 \everypar{\hangindent=\parindent}}

\gdef\refis#1{\indent\hbox to 0pt{\hss#1.~}}    

\gdef\journal#1, #2, #3, #4#5#6#7{               
    {\sl #1~}{\bf #2}, #3 (#4#5#6#7)}           

\def\refstylenp{                
  \gdef\refto##1{ [##1]}                                
  \gdef\refis##1{\indent\hbox to 0pt{\hss##1)~}}        
  \gdef\journal##1, ##2, ##3, ##4 {                     
     {\sl ##1~}{\bf ##2~}(##3) ##4 }}

\def\refstyleprnp{              
  \gdef\refto##1{ [##1]}                                
  \gdef\refis##1{\indent\hbox to 0pt{\hss##1)~}}        
  \gdef\journal##1, ##2, ##3, 1##4##5##6{               
    {\sl ##1~}{\bf ##2~}(1##4##5##6) ##3}}

\def\prd{\journal Phys. Rev. D, }

\def\endreferences{\body}

\def\figurecaptions             
  { \beginparmode
   \subhead{Figure Captions}
}

\def\endpage                    
  {\vfill\eject}

\def\endpaper                   
  {\endmode\vfill\supereject}

\def\hook{\mathbin{\raise2.5pt\hbox{\hbox{{\vbox{\hrule height.4pt 
width6pt depth0pt}}}\vrule height3pt width.4pt depth0pt}\,}}
\def\today{\number\day\ \ifcase\month\or
     January\or February\or March\or April\or May\or June\or
     July\or August\or September\or October\or November\or
     December\space \fi\ \number\year}
\def\date{\noindent{\tt 
     Date typeset: \today\par\bigskip}}
\def\ref#1{Ref. #1}                     
\def\Ref#1{Ref. #1}                     

\def\frac#1#2{{\textstyle{#1 \over #2}}}
\def\half{{\textstyle{ 1\over 2}}}
\def\>{\rangle}
\def\<{\langle}
\def\eg{{\it e.g.,\ }}

\def\ie{{\it i.e.,\ }}

\def\etc{{\it etc.}}

\def\sla{\raise.15ex\hbox{$/$}\kern-.57em}
\def\leaderfill{\leaders\hbox to 1em{\hss.\hss}\hfill}
\def\twiddle{\lower.9ex\rlap{$\kern-.1em\scriptstyle\sim$}}
\def\bigtwiddle{\lower1.ex\rlap{$\sim$}}
\def\gtwid{
\mathrel{\raise.3ex\hbox{$>$\kern-.75em\lower1ex\hbox{$\sim$}}}}
\def\ltwid{\mathrel{\raise.3ex\hbox
{$<$\kern-.75em\lower1ex\hbox{$\sim$}}}}
\def\square{\kern1pt\vbox{\hrule height 1.2pt\hbox
{\vrule width 1.2pt\hskip 3pt
   \vbox{\vskip 6pt}\hskip 3pt\vrule width 0.6pt}
\hrule height 0.6pt}\kern1pt}

\def\m@th{\mathsurround=0pt }
\def\leftrightarrowfill{$\m@th \mathord\leftarrow \mkern-6mu
 \cleaders\hbox{$\mkern-2mu \mathord- \mkern-2mu$}\hfill
 \mkern-6mu \mathord\rightarrow$}
\def\overleftrightarrow#1{\vbox{\ialign{##\crcr
     \leftrightarrowfill\crcr\noalign{\kern-1pt\nointerlineskip}
     $\hfil\displaystyle{#1}\hfil$\crcr}}}


\font\titlefont=cmr10 scaled\magstep3

\def\martinstyletitle                      
  {\null\vskip 3pt plus 0.2fill
   \beginlinemode \doublespace \raggedcenter \titlefont}

\font\twelvesc=cmcsc10 scaled 1200

\def\author                     
  {\vskip 3pt plus 0.2fill \beginlinemode
   \singlespace \raggedcenter\twelvesc}


\def\heading                            
  {\vskip 0.5truein plus 0.1truein      
\endheading
   \beginparmode \def\\{\par} \parskip=0pt \singlespace \raggedcenter}

\def\endheading
  {\par\nobreak\vskip 0.25truein\nobreak\beginparmode}

\def\subheading                         
  {\vskip 0.25truein plus 0.1truein     
   \beginlinemode \singlespace \parskip=0pt \def\\{\par}\raggedcenter}

\def\tag#1$${\eqno(#1)$$}

\def\align#1$${\eqalign{#1}$$}

\def\aligntag#1$${\gdef\tag##1\\{&(##1)\cr}\eqalignno{#1\\}$$
  \gdef\tag##1$${\eqno(##1)$$}}

\def\endaligntag{}

\def\overset #1\to#2{{\mathop{#2}\limits^{#1}}}
\def\underset#1\to#2{{\let\next=#1\mathpalette\undersetpalette#2}}
\def\undersetpalette#1#2{\vtop{\baselineskip0pt
\ialign{$\mathsurround=0pt #1\hfil##\hfil$\crcr#2\crcr\next\crcr}}}


\def\ref#1{Ref.~#1}                     
\def\Ref#1{Ref.~#1}                     
\def\[#1]{[\cite{#1}]}
\def\cite#1{[#1]}
\def\(#1){(\call{#1})}
\def\call#1{{#1}}
\def\taghead#1{}
\def\frac#1#2{{#1 \over #2}}
\def\half{{\frac 12}}

\def\12{{1\over2}}
\def\eg{{\it e.g.,\ }}

\def\ie{{\it i.e.,\ }}

\def\etc{{\it etc.\ }}

\def\sla{\raise.15ex\hbox{$/$}\kern-.57em}
\def\leaderfill{\leaders\hbox to 1em{\hss.\hss}\hfill}
\def\twiddle{\lower.9ex\rlap{$\kern-.1em\scriptstyle\sim$}}
\def\bigtwiddle{\lower1.ex\rlap{$\sim$}}
\def\gtwid{\mathrel{\raise.3ex\hbox{$>$
\kern-.75em\lower1ex\hbox{$\sim$}}}}
\def\ltwid{\mathrel{\raise.3ex\hbox{$<$
\kern-.75em\lower1ex\hbox{$\sim$}}}}
\def\square{\kern1pt\vbox{\hrule height 1.2pt\hbox
{\vrule width 1.2pt\hskip 3pt
   \vbox{\vskip 6pt}\hskip 3pt\vrule width 0.6pt}
\hrule height 0.6pt}\kern1pt}
\def\tdot#1{\mathord{\mathop{#1}\limits^{\kern2pt\ldots}}}

\def\pmb#1{\setbox0=\hbox{#1}%
  \kern-.025em\copy0\kern-\wd0
  \kern  .05em\copy0\kern-\wd0
  \kern-.025em\raise.0433em\box0 }

\catcode`@=11
\newcount\tagnumber\tagnumber=0

\immediate\newwrite\eqnfile
\newif\if@qnfile\@qnfilefalse
\def\write@qn#1{}
\def\writenew@qn#1{}
\def\w@rnwrite#1{\write@qn{#1}\message{#1}}
\def\@rrwrite#1{\write@qn{#1}\errmessage{#1}}

\def\taghead#1{\gdef\t@ghead{#1}\global\tagnumber=0}
\def\t@ghead{}

\expandafter\def\csname @qnnum-3\endcsname
  {{\t@ghead\advance\tagnumber by -3\relax\number\tagnumber}}
\expandafter\def\csname @qnnum-2\endcsname
  {{\t@ghead\advance\tagnumber by -2\relax\number\tagnumber}}
\expandafter\def\csname @qnnum-1\endcsname
  {{\t@ghead\advance\tagnumber by -1\relax\number\tagnumber}}
\expandafter\def\csname @qnnum0\endcsname
  {\t@ghead\number\tagnumber}
\expandafter\def\csname @qnnum+1\endcsname
  {{\t@ghead\advance\tagnumber by 1\relax\number\tagnumber}}
\expandafter\def\csname @qnnum+2\endcsname
  {{\t@ghead\advance\tagnumber by 2\relax\number\tagnumber}}
\expandafter\def\csname @qnnum+3\endcsname
  {{\t@ghead\advance\tagnumber by 3\relax\number\tagnumber}}

\def\equationfile{%
  \@qnfiletrue\immediate\openout\eqnfile=\jobname.eqn%
  \def\write@qn##1{\if@qnfile\immediate\write\eqnfile{##1}\fi}
  \def\writenew@qn##1{\if@qnfile\immediate\write\eqnfile
    {\noexpand\tag{##1} = (\t@ghead\number\tagnumber)}\fi}
}

\def\callall#1{\xdef#1##1{#1{\noexpand\call{##1}}}}
\def\call#1{\each@rg\callr@nge{#1}}

\def\each@rg#1#2{{\let\thecsname=#1\expandafter\first@rg#2,\end,}}
\def\first@rg#1,{\thecsname{#1}\apply@rg}
\def\apply@rg#1,{\ifx\end#1\let\next=\relax%
\else,\thecsname{#1}\let\next=\apply@rg\fi\next}

\def\callr@nge#1{\calldor@nge#1-\end-}
\def\callr@ngeat#1\end-{#1}
\def\calldor@nge#1-#2-{\ifx\end#2\@qneatspace#1 %
  \else\calll@@p{#1}{#2}\callr@ngeat\fi}
\def\calll@@p#1#2{\ifnum#1>#2{\@rrwrite
{Equation range #1-#2\space is bad.}
\errhelp{If you call a series of equations by the notation M-N, then M and
N must be integers, and N must be greater than or equal to M.}}\else %
{\count0=#1\count1=
#2\advance\count1 by1\relax\expandafter\@qncall\the\count0,%
  \loop\advance\count0 by1\relax%
    \ifnum\count0<\count1,\expandafter\@qncall\the\count0,%
  \repeat}\fi}

\def\@qneatspace#1#2 {\@qncall#1#2,}
\def\@qncall#1,{\ifunc@lled{#1}{\def\next{#1}\ifx\next\empty\else
  \w@rnwrite{Equation number \noexpand\(>>#1<<) 
has not been defined yet.}
  >>#1<<\fi}\else\csname @qnnum#1\endcsname\fi}

\let\eqnono=\eqno
\def\eqno(#1){\tag#1}
\def\tag#1$${\eqnono(\displayt@g#1 )$$}

\def\aligntag#1\endaligntag
  $${\gdef\tag##1\\{&(##1 )\cr}\eqalignno{#1\\}$$
  \gdef\tag##1$${\eqnono(\displayt@g##1 )$$}}

\def\eqalignno#1{\displ@y \tabskip\centering
  \halign to\displaywidth{\hfil$\displaystyle{##}$\tabskip\z@skip
    &$\displaystyle{{}##}$\hfil\tabskip\centering
    &\llap{$\displayt@gpar##$}\tabskip\z@skip\crcr
    #1\crcr}}

\def\displayt@gpar(#1){(\displayt@g#1 )}

\def\displayt@g#1 {\rm\ifunc@lled{#1}\global\advance\tagnumber by1
        {\def\next{#1}\ifx\next\empty\else\expandafter
        \xdef\csname
 @qnnum#1\endcsname{\t@ghead\number\tagnumber}\fi}%
  \writenew@qn{#1}\t@ghead\number\tagnumber\else
        {\edef\next{\t@ghead\number\tagnumber}%
        \expandafter\ifx\csname @qnnum#1\endcsname\next\else
        \w@rnwrite{Equation \noexpand\tag{#1} is 
a duplicate number.}\fi}%
  \csname @qnnum#1\endcsname\fi}

\def\ifunc@lled#1{\expandafter\ifx\csname @qnnum#1\endcsname\relax}

\let\@qnend=\end\gdef\end{\if@qnfile
\immediate\write16{Equation numbers 
written on []\jobname.EQN.}\fi\@qnend}

\catcode`@=12

\catcode`@=11
\newcount\r@fcount \r@fcount=0
\newcount\r@fcurr
\immediate\newwrite\reffile
\newif\ifr@ffile\r@ffilefalse
\def\w@rnwrite#1{\ifr@ffile\immediate\write\reffile{#1}\fi\message{#1}}

\def\writer@f#1>>{}
\def\referencefile{
  \r@ffiletrue\immediate\openout\reffile=\jobname.ref%
  \def\writer@f##1>>{\ifr@ffile\immediate\write\reffile%
    {\noexpand\refis{##1} = \csname r@fnum##1\endcsname = %
     \expandafter\expandafter\expandafter\strip@t\expandafter%
     \meaning\csname r@ftext
\csname r@fnum##1\endcsname\endcsname}\fi}%
  \def\strip@t##1>>{}}

\def\citeall#1{\xdef#1##1{#1{\noexpand\cite{##1}}}}
\def\cite#1{\each@rg\citer@nge{#1}}	

\def\each@rg#1#2{{\let\thecsname=#1\expandafter\first@rg#2,\end,}}
\def\first@rg#1,{\thecsname{#1}\apply@rg}	
\def\apply@rg#1,{\ifx\end#1\let\next=\relax
\else,\thecsname{#1}\let\next=\apply@rg\fi\next}

\def\citer@nge#1{\citedor@nge#1-\end-}	
\def\citer@ngeat#1\end-{#1}
\def\citedor@nge#1-#2-{\ifx\end#2\r@featspace#1 
  \else\citel@@p{#1}{#2}\citer@ngeat\fi}	
\def\citel@@p#1#2{\ifnum#1>#2{\errmessage{Reference range #1-
#2\space is bad.}%
    \errhelp{If you cite a series of references by the notation M-N, then M 
and
    N must be integers, and N must be greater than or equal to M.}}\else%
 {\count0=#1\count1=#2\advance\count1 
by1\relax\expandafter\r@fcite\the\count0,
  \loop\advance\count0 by1\relax
    \ifnum\count0<\count1,\expandafter\r@fcite\the\count0,%
  \repeat}\fi}

\def\r@featspace#1#2 {\r@fcite#1#2,}	
\def\r@fcite#1,{\ifuncit@d{#1}
    \newr@f{#1}%
    \expandafter\gdef\csname r@ftext\number\r@fcount\endcsname%
                     {\message{Reference #1 to be supplied.}%
                      \writer@f#1>>#1 to be supplied.\par}%
 \fi%
 \csname r@fnum#1\endcsname}
\def\ifuncit@d#1{\expandafter\ifx\csname r@fnum#1\endcsname\relax}%
\def\newr@f#1{\global\advance\r@fcount by1%
    \expandafter\xdef\csname r@fnum#1\endcsname{\number\r@fcount}}

\let\r@fis=\refis			
\def\refis#1#2#3\par{\ifuncit@d{#1}
   \newr@f{#1}%
   \w@rnwrite{Reference #1=\number\r@fcount\space is not cited up to
 now.}\fi%
  \expandafter
\gdef\csname r@ftext\csname r@fnum#1\endcsname\endcsname%
  {\writer@f#1>>#2#3\par}}

\def\ignoreuncited{
   \def\refis##1##2##3\par{\ifuncit@d{##1}%
    \else\expandafter\gdef
\csname r@ftext\csname r@fnum##1\endcsname\endcsname%
     {\writer@f##1>>##2##3\par}\fi}}

\def\r@ferr{\endreferences\errmessage{I was expecting to see
\noexpand\endreferences before now;  I have inserted it here.}}
\let\r@ferences=\references
\def\references{\r@ferences\def\endmode{\r@ferr\par\endgroup}}

\let\endr@ferences=\endreferences
\def\endreferences{\r@fcurr=0
  {\loop\ifnum\r@fcurr<\r@fcount
    \advance\r@fcurr by 
1\relax\expandafter\r@fis\expandafter{\number\r@fcurr}%
    \csname r@ftext\number\r@fcurr\endcsname%
  \repeat}\gdef\r@ferr{}\endr@ferences}


\let\r@fend=\endpaper\gdef\endpaper{\ifr@ffile
\immediate\write16{Cross References written on 
[]\jobname.REF.}\fi\r@fend}

\catcode`@=12

\def\reftorange#1#2#3{{\refto{#1}--\setbox0=\hbox{\cite{#2}}\refto{#3}}}

\citeall\refto		
\citeall\ref		%
\citeall\Ref		%

\ignoreuncited
\def\det{{\rm det}}
\def\>{\rangle}
\def\<{\langle}

\def\z{\vec z}
\def\zetabar{\bar \zeta}
\def\veczeta{\vec \zeta}
\def\veczetaprime{\vec{\zeta}^{\,\prime}}

\def\Sym{{\cal S}}
\def\Tr{{\rm Tr}}
\def\tr{{\rm tr}}
\def\V{{\cal V}}
\def\H{{\cal H}}

\def\w{{\vec w}}
\def\I{{\cal I}}



\line{\hfill \tenrm March 2005}
\bigskip\bigskip
\centerline{\bigbf ON THE COHERENT STATE PATH INTEGRAL}
\centerline{\bigbf  FOR LINEAR SYSTEMS}
\bigskip
\centerline{C. G. Torre}
\centerline{\sl Department of Physics, Utah State University, Logan, Utah, 84322-4415 USA}
\bigskip

\singlespace
{\tenrm
We present a computation of the coherent state path integral for a generic linear system using ``functional methods'' (as opposed to discrete time approaches). The Gaussian phase space path integral is formally given by a determinant built from a first-order differential operator with coherent state boundary conditions. We show how this determinant can be expressed in terms of the symplectic transformation generated by the (in general, time-dependent) quadratic Hamiltonian for the system.  We briefly discuss the conditions under which the coherent state path integral for a linear system actually exists. A necessary --- but not sufficient --- condition for existence of the path integral is that the symplectic transformation generated by the Hamiltonian is (unitarily) implementable on the Fock space for the system.}

\oneandahalfspace
\taghead{1.}
\head{1. Introduction}

The coherent state path integral, a variant of the phase space path integral, has long been recognized as a useful tool in quantum mechanics and in quantum field theory (see, \eg \reftorange{Itzykson1980} {Faddeev1980,Schulman1981,Klauder1985} {Baranger2001}). In the quantum mechanical setting this form of the path integral has been studied fairly extensively (see, \eg \refto{ Klauder1985}), although it would seem that only relatively recently  does there appear a definitive treatment of the coherent state path integral for a generic linear system with one degree of freedom (in the context of a study of the semi-classical approximation) \refto{Baranger2001}. In quantum mechanics the method normally used to define and analyze the coherent state path integral is based upon taking a limit of an approximation based upon discretized paths. In the field theoretic setting the coherent state path integral is mainly used to set up the perturbative evaluation of the S-matrix; here ``functional methods'' (as opposed to discretized path methods) are normally employed. In particular, at the level of the free or semi-classical theory the path integral representation of the vacuum to vacuum transition amplitude, after integrating out the canonical momenta, is expressed in terms of a Fredholm determinant of a linear differential operator (wave operator, Dirac operator, \etc), which features in the approximate quadratic action functional.  

In this paper we compute the coherent state path integral for a generic linear bosonic dynamical system with general coherent state boundary conditions.  The number of degrees of freedom can be infinite, so this computation includes field theory.  In our computation we do not first integrate out the canonical momenta, nor do we restrict attention to relativistic fields. Therefore, the path integral is formally given in terms of the determinant of a first-order differential operator with coherent state boundary conditions. We show how this determinant can be expressed in terms of the symplectic transformation generated by the (in general, time dependent) quadratic Hamiltonian for the system. While our computations are somewhat formal, the resulting expression for the coherent state transition amplitude agrees with the (rigorous) result  obtained using methods of canonical quantization \refto{Honegger1996}, as it should.

The computation provided here demonstrates a viable method for evaluating a class of coherent state path integrals which we hope will be a useful addition to the collection of techniques used to compute path integrals.  This computation should be relevant for a number of applications, including: any linear quantum mechanical system, linearized/semi-classical approximations,  quantum fields in curved spacetime, and parametrized free field theory and various other quantum gravity models \refto{MV2004}.

Our results are also intended to help explore within the path integral formalism subtle quantum field theoretic phenomena that have been uncovered using other methods of quantization (\eg canonical quantization). In particular, it is well-known that for field theories there exist inequivalent representations of the canonical commutation relations (see, \eg \refto{Bratteli1979} and references therein). Closely related to this is the fact that many linear canonical transformations --- which may include those defining the time evolution of a linear system --- cannot be unitarily implemented in the Fock space quantization of a field theory \refto{Shale1962, Berezin1966, Honegger1996}.  This situation is known to occur in a variety of physical settings, \eg for quantum fields in curved spacetimes \refto{Helfer1996}, for the polarized Gowdy model in General Relativity \refto{Gowdy}, and for parametrized free field theories in dimensions greater than two \refto{CGT1999c}.  It is natural to ask how these important phenomena, which have been understood heretofore using operator techniques, manifest themselves in the path integral formalism.  Because we can express the  path integral  in terms of the symplectic transformations generated by the classical Hamiltonian, the connection with results from canonical quantization on unitary implementability/equivalence becomes accessible. For example, in this paper we shall see explicitly that unitary implementability 
of dynamical evolution is necessary but not sufficient for the coherent state path integral to exist.

\taghead{2.}
\head{2. Preliminaries}

We will be considering the path integral for a linear dynamical system, by which we mean the following. Fix a real Hilbert space, \ie a real vector space $\V$, complete  with respect to a scalar product $ (\cdot ,\cdot )$. Elements of $\V$ will be denoted $\z$, $\w$, \etc The vector space $\V$ is to be the phase space for the system, so we further assume that $\V$ is equipped with a densely defined symplectic form $\Omega$ and a Hamiltonian $\H$, which is a densely defined quadratic form on $\V$. As explained \eg in \refto{Wald1994}, we require the inner product and symplectic form satisfy
$$
(\z,\z) = {1\over 4}\, {{\rm l.u.b.}\atop \w\neq0} {[\Omega(\z,\w)]^2\over (\w,\w)}.
\tag
$$
This implies that the symplectic form is bounded on $\V$ and so can be defined with $\V$ as its domain.  The scalar product and symplectic form then combine to define a (bounded, skew-adjoint) complex structure $J\colon \V\to \V$ via
$$
\Omega(\z,\w) = 2(\z,J\w),
\tag
$$
 which can be used to define ``positive and negative frequency'' solutions to the linear field equations defined by $\H$. More precisely, we introduce the complexification $\V^C$ of $\V$ and extend $ (\cdot ,\cdot )$, $\Omega$ and $J$ to $\V^C$ via linearity.    We define complex conjugate Hilbert spaces $\V_\pm\subset \V^C$ corresponding to the $\pm i$ eigenspaces of $J$ equipped with the sesquilinear inner products 
 $$
 \z,\w\in \V_\pm \to 2({\z}^{\,*},\w).
 \tag
 $$ 
  Here we use an asterisk for complex conjugation  on $\V^C$.
For later convenience, we suppose we have fixed a basis: $\{\vec e_a\}$ for $\V_+$ and  $\{\vec e_a^{\,*}\}$ for $\V_-$. If $\vec z\in \V$, we have\footnote*{The summation convention is in effect and the range of the summation can be infinite and continuous (\ie integration).  }
$$
\vec z = z^a\vec e_a + z^{a*}\vec e_a^{\,*},
\tag
$$
a relationship we will often denote as either
$$
\vec z = \left(\matrix{z^{a\phantom*}\cr  z^{a*}}\right),
\tag
$$
or simply
$$
\vec z= \left(\matrix{z^{\phantom{*}}\cr  z^{*}}\right).
\tag
$$
If $\vec z\in \V^C$ we write
$$
\vec z = z^a\vec e_a + \bar z^{a}\vec e_a^{\,*},
\tag
$$
or
$$
\vec z = \left(\matrix{z^a\cr \bar z^a}\right),
\tag vecz
$$
or simply
$$
\vec z =\left(\matrix{z\cr  \bar z}\right),
\tag
$$
so that $\vec z = \left(\matrix{z^a\cr \bar z^a}\right)$ is in fact an element of $\V$ if and only if $\bar z^a = z^{a*}$.  With this notation we have
$$
\vec z \in \V_+ \quad\longleftrightarrow\quad \vec z= z^a \vec e_a \quad\longleftrightarrow\quad \vec z = \left(\matrix{z^a\cr 0}\right),
\tag
$$
$$
\vec z \in \V_- \quad\longleftrightarrow\quad \vec z= \bar z^a \vec e_a^{\,*} \quad\longleftrightarrow\quad \vec z = \left(\matrix{0\cr \bar z^a}\right),
\tag
$$
$$
\vec z^{\, *} \in \V_- \quad\longleftrightarrow\quad \vec z= z^{a*} \vec e_a^{\,*} \quad\longleftrightarrow\quad \vec z = \left(\matrix{0\cr z^{a*}}\right).
\tag
$$
The basis $\{\vec e_a, \vec e_a^{\,*}\}$ is chosen such that
$$
(\vec w,\vec z) 
=\half(\bar w_a z^a + w^a \bar z_a),
\tag
$$
$$
\Omega(\vec w,\vec z) = {1\over i}(\bar z_a w^a - \bar w_a z^a),
\tag cansym
$$
$$
J\vec z = \left(\matrix{\phantom{-} i z^a\cr - i \bar z^a}\right),
\tag
$$
and
$$
\H(\z) = A_{ab}  z^a\bar z^b + \half B_{ab}z^a z^b + \half\bar B_{ab}\bar z^a\bar z^b.
\tag Ham
$$
where $A$ is self-adjoint and $B$, $\bar B=B^*$ are symmetric.

As explained, \eg in \refto{Wald1994} the data $\V$, $\Omega$, $\mu$ define a ``1-particle'' Hilbert space and a corresponding  Fock space representation of the Heisenberg group. In this representation, the operators corresponding to the classical variables $z^b$ and $\bar z^b$ are annihilation and  creation operators $a^b$ and $a^{\dagger b}$, respectively.

The path integral we will be studying is over a space of paths $\vec z = \vec z(t)$ in $\V^C$. The paths will be required to obey coherent state boundary conditions of the form
$$
z^a(0) = w^a,\quad \bar z^a(T) =v^a,
\tag Ibc
$$
for some given  $w^a$ and $v^a$.
Given the Hamiltonian \(Ham), symplectic structure \(cansym), and boundary conditions \(Ibc)  the action functional on this space of paths is given by
$$
\eqalign{
Q(\vec z) =\int_0^T dt\, \Bigg\{&{1\over 2i} (\dot{\bar z}_a z^a - \bar z_a \dot z^a)
- (A_{ab}  z^a\bar z^b + \half B_{ab}z^a z^b + \half\bar B_{ab}\bar z^a\bar z^b)\Bigg\}\cr
&\quad +{1\over 2i}\big[\bar z_a(T) z^a(T) + \bar z_a(0) z^a(0)\big].
}
\tag Q
$$
This is just the phase space action functional on $\V$ expressed in a complex coordinate chart and extended to $\V^C$.
  The boundary terms in \(Q) are there so that $Q$ is differentiable with the boundary conditions \(Ibc) \refto{Faddeev1980}. 
 
In light of \(Ibc), we fix a path $\vec{\bf z}(t)$ with these boundary conditions and set
$$
\z(t) = \vec{\bf z}(t) + \vec\zeta(t).
\tag
$$ 
The variables $\vec\zeta(t)$ satisfy the boundary conditions
$$
\zeta^a(0) = 0,\quad \bar\zeta^a(T) = 0,
\tag vacbc
$$
and are elements of a vector space $X$ with dual $X^\prime$. The pairing between $X$ and $X^\prime$ is
$$
\veczeta(t) =\left(\matrix{\zeta^a\cr \bar\zeta^a}\right)\in X,\quad 
\veczetaprime(t)=\left(\matrix{\zeta^{\prime}_a\cr \bar\zeta^{\prime}_a}\right)\in X^\prime\  \longrightarrow\  \<\veczetaprime,\veczeta\>=\int_0^T dt\, \half\Big(\bar \zeta^{\, \prime}_a(t) \zeta^a(t) + \zeta^{\, \prime}_a(t)\bar \zeta^a(t)\Big).
\tag
$$ 
The action restricted to $X$ is a quadratic form characterized by a symmetric linear operator $D\colon X\to X^\prime$:
$$
Q(\vec \zeta) = \<D\vec \zeta,\vec \zeta\>,
\tag squad
$$
where 
$$
D\veczeta = \left(\matrix{i\dot\zeta_a - A_{ba}\zeta^b - \bar B_{ab}\bar\zeta^b\cr
-i\dot{\bar\zeta}_a - A_{ab}\bar\zeta^b - B_{ab}\zeta^b
}\right)
.
\tag
$$
We also define the quadratic forms $Q_0$ and $V$  on $X$ and symmetric operators $D_0\colon X\to X^\prime$ and $N\colon X\to X^\prime$ by
 $$
 \eqalign{
 Q_0(\veczeta) &= \int_0^T dt\, \Bigg\{{1\over 2i} (\dot{\zetabar}_a \zeta^a - \zetabar_a \dot \zeta^a)
- A_a^b \zetabar_b \zeta^a\Bigg\} 
\cr
&=\<D_0\vec \zeta,\vec \zeta\> ,
}
\tag
$$
$$
D_0\veczeta = \left(\matrix{i\dot\zeta_a - A_{ba}\zeta^b\cr
-i\dot{\bar\zeta}_a - A_{ab}\bar\zeta^b
}\right)
,
\tag
$$
and
$$
Q(\veczeta) = Q_0(\veczeta) + V(\veczeta),
\tag
$$
where
$$
V(\veczeta) = \<N \veczeta,\veczeta\>,\quad N\veczeta = \left(\matrix{
- \bar B_{ab}\bar\zeta^b\cr
-B_{ab}\zeta^b
}\right).
\tag ndef
$$


\taghead{3.}
\head{3. The Path Integral}

Our goal is to compute a path integral which is, roughly, of the form
$$
{\cal I}(v,w) \sim \int\, [d\vec z] e^{i  Q(\vec z)},
\tag fpi
$$
where the integral is over a space of phase space paths with the boundary conditions \(Ibc). 
From the point of view of the Fock space representation of the quantum theory, $\I(v,w)$ should be a coherent state matrix element of the time evolution operator $U(0,T)$ defined by a Hamiltonian corresponding to the classical observable $\H$:
$$
{\cal I}(v,w) = \<v|U(0,T)|w\>.
\tag
$$
Here the states $|v\>$ and $|w\>$ are Fock space eigenvectors of the annihilation operator $a^b$:
$$
a^b |v\> = v^b|v\>,\quad a^b|w\> = w^b|w\>.
\tag
$$

An approach to obtaining a rigorous definition of \(fpi) can be  found in \refto{Albeverio1976, DeWitt1995}.  We shall not attempt to define functional integration here; instead we shall proceed formally by postulating a few basic properties, which feature in the rigorous definitions of \refto{Albeverio1976, DeWitt1995}, and which any suitable definition of integration should exhibit.   We shall assume that integration is a linear operation on a class of functions on the vector space $X$. The integration operation is normalized relative to the quadratic form $Q_0$ and is denoted with the symbol
$
\int_X [d\zeta]_{Q_0}.
$ 
The normalization condition is
$$
\int_X [d\veczeta]_{Q_0} e^{i Q_0(\veczeta)} = 1.
\tag fpinorm
$$
We formally define the path integral \(fpi) by
$$
{\cal I}(v,w) = \int_X [d\veczeta]_{Q_0}\, e^{i Q(\vec{\bf z} + \veczeta)}.
\tag
$$
As we shall see, in the Fock space representation our choice of path integral normalization corresponds to defining the time evolution operator using the normal-ordered Hamiltonian operator associated to the classical expression \(Ham), with no additive $c$-number renormalizations.

To evaluate the path integral we choose 
$$
\vec{\bf z}(t)=\left(\matrix{{\bf z}(t)\cr\bar{\bf z}(t)}\right)
\tag
$$
 to be a critical point of the action functional, \ie  $\vec {\bf z}(t)$ is chosen to be the (unique) path in $\V^C$ satisfying 
$$
D\vec{\bf z}(t) = 0, \quad {\bf z}(0) = w,\ \bar{\bf z}(T) = v.
\tag
$$
 Using linearity of the integration operation,
the path integral now takes the form
$$
\I(v,w) = \exp\left\{{1\over 2}\big[v_a {\bf z}^a(T) + \bar{\bf  z}_a(0) w^a\big]\right\}\int_X [d\veczeta]_{Q_0}e^{i Q(\veczeta)}.
\tag shifti
$$
The remaining path integral is an oscillating Gaussian --- or Fresnel -- type of integral. It is (at least formally) given in terms of the Fredholm determinant  \refto{DeWitt1995}:\footnote*{
In principle there could also be a phase factor coming from the index of $Q$ relative to $Q_0$ \refto{DeWitt1995}, but we shall see that this factor is unity.}
$$
{\cal I}(0,0)= \int_X [d\veczeta]_{Q_0}e^{i Q(\veczeta)}
= \det^{-\half}(D_0^{-1}D).
\tag
$$
Hence
$$
\I(v,w) = \exp\left\{{1\over 2}\big[v_a {\bf z}^a(T) + \bar{\bf  z}_a(0) w^a\big]\right\}\det^{-\half}(D_0^{-1}D).
\tag prelimans
$$

\taghead{4.}
\head{4. Evaluating the path integral in terms of classical dynamics}

We now show that the result \(prelimans) can be expressed in terms of the symplectic transformation(s) generated by ${\cal H}(t)$ and the boundary conditions \(Ibc). We begin by  giving some relevant features of the symplectic transformation. 

We assume the Hamiltonian $\H(t)$, which is allowed to be time-dependent,  generates a 1-parameter family of symplectic transformations $\Sym(t)$ on $\V^C$. $\Sym$ is determined by
$$
\left({d\over dt} + H(t)\right) \Sym(t) = 0,\quad H(t)  = i\left(\matrix{
A(t) & \bar B(t)\cr -  B(t) & -A(t)
}\right),\quad \Sym(0)=id.
\tag seq
$$
The operators $\Sym(t)$ can be expressed as a Dyson-type of expansion, \ie the time-ordered exponential of the linear transformations $H(t)$ on $\V^C$ defined by the Hamiltonian $\H(t)$. We shall not need the explicit formula here.
Using a block matrix notation paralleling
\(vecz), we can express the symplectic transformations as
$$
\Sym(t) = \left(\matrix{\alpha(t) &\beta(t)\cr \bar\beta(t) &\bar \alpha(t)}
\right),
\tag Salphabeta
$$
\ie 
$$
\Sym{\bf \z}=\left(\matrix{ \alpha^a_b {\bf z}^b + \beta^{a}_{b}\bar {\bf z}^b,\cr
\bar\alpha_b^a \bar {\bf z}^b + \bar\beta^{a}_b{\bf z}^b}\right),
\tag
$$
where the Bogoliubov coefficients $\alpha(t)$, $\beta(t)$  satisfy at each $t$
$$
\alpha\colon \V_+\to \V_+,\quad \beta\colon \V_-\to\V_+,
\tag bc1
$$
$$
\bar\alpha\colon \V_-\to\V_-,\quad \bar\beta\colon \V_+\to\V_-,
\tag bc2
$$
$$
\bar\alpha=\alpha^*,\quad \bar\beta=\beta^*,\quad
\tag bc3
$$
$$
\alpha\alpha^\dagger - \beta\beta^\dagger = id_{\V_+},\quad \alpha\beta^T - \beta\alpha^T = 0.
\tag canid
$$
 From these equations it follows that
$\alpha(t)$ has a bounded inverse and that
$$
\Sym^{-1}(t) = \left(\matrix{\alpha^{\dagger}(t) &-\beta^T(t)\cr
-\beta^\dagger(t) &\alpha^T(t)}\right).
\tag Sinvalphabeta
$$
Finally, we have that
$$
\alpha(0) = id_{\V_+},\quad \bar\alpha(0) = id_{\V_-},\quad \beta(0) =0,\quad \bar\beta(0) = 0.
\tag bc4
$$

 All the dependence of $\I(v,w)$ on the initial and final states is in the exponential of the action sitting in front of the determinant in \(prelimans).   To make this initial/final state dependence explicit, we use the fact that the critical points 
 $$
 \vec{\bf z}(t) = \left(\matrix{{\bf z}(t)\cr\bar{\bf z}(t)}\right)
 \tag
 $$ 
 of the action are determined by the symplectic transformations generated by the Hamiltonian \(Ham):
$$
\vec {\bf z}(t) = \Sym(t) \vec {\bf z}(0).
\tag
$$
The boundary conditions \(Ibc) imply
$$
{\bf z}(T) = \alpha^{-1\dagger}(T) w + \sigma(T) v
\tag
$$
$$
\bar{\bf z}(0) = \bar\alpha^{-1}(T)\, v -\gamma(T) w,
\tag
$$
where we have defined symmetric operators $\gamma\colon \V_+\to \V_-$ and $\sigma\colon \V_-\to \V_+$ via 
$$
\gamma = \bar\alpha^{-1}\bar\beta, \quad \sigma = \beta \bar\alpha^{-1}.
\tag
$$

Putting this all together, we get
\def\expiS{\exp\left\{(\alpha^{-1\dagger})^a_b(T)v_aw^b + \half \sigma^{a}_{b}(T)v_av^b - \half \gamma_{a}^{b}(T) w^a w_b\right\}}
$$
\I(v,w) = \expiS\det^{-\half}(D_0^{-1}D).
\tag theintegral
$$

The determinant appearing in \(theintegral) depends upon the Hamiltonian \(Ham) via the symplectic transformation $\Sym$.  We will now make this dependence on $\Sym$ explicit.   The hypothesis on $M=D_0^{-1}D$ is that $M-1$ is trace class; this guarantees the determinant is well defined and satisfies the variational identity.$$
\delta \log \det (M) = {\rm Tr}( M^{-1}\delta M).
\tag varid
$$
The trace (``Tr'') of an operator $R\colon X\to X$ is defined in terms of an integral kernel $R(t,u)$ with values in the set of operators on $\V^C$:
$$
R\, \vec z(t) = \int_0^T du\, R(t,u) \vec z(u).
\tag
$$
We set
$$
{\rm Tr}(R) = \int_0^T dt\, {\rm tr} R(t,t),
\tag
$$
where ``tr'' denotes the Hilbert space trace. 
To use \(varid) we define the differential operator $D_\lambda\colon X\to X^\prime$ by
$$
D_\lambda\veczeta = \left(\matrix{i\dot\zeta_a - A_{ba}\zeta^b - \lambda \bar B_{ab}\bar\zeta^b\cr
-i\dot{\bar\zeta}_a - A_{ab}\bar\zeta^b -\lambda B_{ab}\zeta^b
}\right)
,
\tag
$$
(so that $D = D_1$). 
Setting
$$
\I_\lambda = \det^{-\half}(D_0^{-1}D_\lambda),
\tag
$$
we have
$$
{d\over d\lambda}\log \I_\lambda  = -\half{\rm Tr}\Big\{(D_0^{-1} D_\lambda)^{-1} D_0^{-1}N\Big\} =-\half {\rm Tr}(D_\lambda^{-1} N),
\tag nude
$$
where $N$ was defined in \(ndef).
Our strategy is to obtain a suitable expression of ${\rm Tr}(D_\lambda^{-1} N)$ in terms of symplectic transformations and then solve the differential equation \(nude) in $\lambda$  with initial condition given by the path integral normalization,
$$
{\cal I}_0 = 1,
\tag ibc
$$
to find ${\cal I}_\lambda$, from which we have ${\cal I} = {\cal I}_{\lambda=1}$.\footnote*{This approach will define $\cal I$ provided $D_\lambda$ has no zero eigenvalues as $\lambda$ varies from $0$ to $1$ \refto{DeWitt1995}. Otherwise there is an additional phase coming from the index of $D_\lambda$. We shall see that $D_\lambda$ has no kernel.}


To obtain a suitable form of ${\rm Tr}(D_\lambda^{-1} N)$ we need to obtain an expression for the Green function $G_\lambda\equiv D_\lambda^{-1}$, which is uniquely determined by the solution of the system
$$
D_\lambda \vec \zeta(t) = \vec F(t),\quad\quad \zeta^a(0) = 0, \quad \bar \zeta^a(T) = 0,
\tag inhomeq
$$
so that
$$
\vec \zeta(t) = G_\lambda\vec F(t)= \int_0^Tdu\, G_\lambda(t,u)\vec F(u).
\tag solinhomeq
$$
The general solution to the differential equation in \(inhomeq) is easily checked to be
$$
\vec \zeta(t) = \vec \zeta_0(t) + \int_0^T du\, \theta(t-u) \Sym_\lambda(t) \Sym_\lambda^{-1}(u)\Sigma\vec F(u),
\tag gensolzt
$$
where $\vec \zeta_0$ is the general solution to $D_\lambda\vec \zeta_0 = 0$, 
$$
\vec \zeta_0(t) = \Sym_\lambda(t)\vec \zeta_0(0),
\tag
$$
with
$$
\Sym_\lambda(t)=\left(\matrix{
\alpha_\lambda(t) &\beta_\lambda(t)\cr
\bar\beta_\lambda(t)&\bar\alpha_\lambda(t)
}\right),
\quad
\Sym_\lambda^{-1}(t) = \left(\matrix{\alpha_\lambda^{\dagger}(t) &-\beta_\lambda^T(t)\cr
-\beta_\lambda^\dagger(t) &\alpha_\lambda^T(t)}\right)
\tag symlambda
$$ 
being the symplectic transformation generated by 
$$
\H_\lambda(t)= A_{ab}(t)  z^a\bar z^b + \half\lambda\left( B_{ab}(t)z^a z^b + \bar B_{ab}(t)\bar z^a\bar z^b\right).
\tag
$$ 
$\Sym_\lambda(t)$ satisfies
$$
\left({d\over dt} + H_\lambda(t)\right) \Sym_\lambda(t) = 0,\quad H_\lambda(t)  = i\left(\matrix{
A(t) &\lambda \bar B(t)\cr -\lambda  B(t) & -A(t)
}\right),
\tag slambdaeq
$$
and, for each value of $\lambda$, the obvious generalizations of equations \(bc1)--\(canid) and \(bc4) hold.
In \(gensolzt) $\theta$ is the step-function
$$
\theta(x) = \cases{1 &$x>0$,\cr
0 &$x<0$,}
\tag
$$
and
$$
\Sigma = \left(\matrix{
-i &0\cr
0& i
}\right).
\tag
$$
The boundary conditions in \(inhomeq) imply (suppressing indices)
$$
\vec \zeta_0(t) = \left(\matrix{\beta_\lambda(t)\bar \zeta_0(0)\cr
\bar\alpha_\lambda(t)\bar \zeta_0(0)}\right),
\tag
$$
where
$$
\bar \zeta_0(0) = -\bar\alpha_\lambda^{-1}(T) \left[ \int_0^T du\, \Sym_\lambda(T) \Sym_\lambda^{-1}(u) \Sigma \vec F(u)\right]_{\bar \zeta}.
\tag
$$
Here we use a notation for the components of elements of $\V^C$ such that if
$$
\vec v = \left(\matrix{v\cr \bar v}\right),
\tag
$$
then
$$
\left(\vec v\right)_{\zeta} = v,\quad \left(\vec v\right)_{\bar \zeta}=\bar v.
\tag
$$
Putting all of this together, the solution
$$
\vec \zeta(t) = \left(\matrix{\zeta(t)\cr \bar \zeta(t)}\right)
\tag
$$
 to \(inhomeq) is given by\footnote\dag{As can be seen from this result, there are no non-trivial solutions to $D_\lambda \vec \zeta = 0$ with boundary conditions \(vacbc). This means that $D_\lambda$ has no kernel and there are no additional phases to be computed (see \refto{DeWitt1995}). Evidently, this is a simplifying feature of the coherent state path integral.}
$$
\eqalign{
\zeta(t) &=\int_0^T du\, \Bigg\{\Big(\theta(t-u)\alpha_\lambda(t) -\beta_\lambda(t)\gamma_\lambda(T)\Big) \left[\Sym_\lambda^{-1}(u)\Sigma\vec F(u)\right]_{\zeta}\cr
&\quad\quad -\theta(u-t)\beta_\lambda(t)\left[\Sym_\lambda^{-1}(u)\Sigma\vec F(u)\right]_{ \bar \zeta}
\Bigg\}\cr
\bar \zeta(t) &=\int_0^T du\, \Bigg\{\Big(\theta(t-u)\bar\beta_\lambda(t) -\bar\alpha_\lambda(t)\gamma_\lambda(T)\Big) \left[\Sym_\lambda^{-1}(u)\Sigma\vec F(u)\right]_{\zeta}\cr
&\quad\quad -\theta(u-t)\bar\alpha_\lambda(t)\left[\Sym_\lambda^{-1}(u)\Sigma\vec F(u)\right]_{ \bar \zeta}
\Bigg\},
}
\tag
$$
where
$$
\left[\Sym_\lambda^{-1}\Sigma \vec F\right]_{\zeta} = -i\alpha_\lambda^\dagger F_\zeta -i \beta_\lambda^T F_{\bar \zeta},
\quad
\left[\Sym_\lambda^{-1}\Sigma \vec F\right]_{\bar \zeta} = i\beta_\lambda^\dagger F_\zeta + i\alpha_\lambda^T F_{\bar \zeta}.
\tag
$$
%
%
The Green function thus takes the form
$$
G_\lambda(t,u)=\left(\matrix{G_{\zeta\zeta}(t,u) &G_{\zeta\bar \zeta}(t,u)\cr
G_{\bar \zeta \zeta}(t,u) & G_{\bar \zeta\bar \zeta}(t,u)}\right),
\tag
$$
where
$$
\eqalign{
G_{\zeta\zeta}(t,u) &= -i\theta(t-u)\bigg(\alpha_\lambda(t)\alpha^\dagger_\lambda(u) -\beta_\lambda(t)\beta_\lambda^\dagger(u)\bigg)
 +i\beta_\lambda(t)\bigg(\gamma_\lambda(T)\alpha_\lambda^\dagger(u) -\beta_\lambda^\dagger(u)\bigg),
\cr
G_{\zeta\bar \zeta}(t,u) &= -i\theta(t-u) \bigg(\alpha_\lambda(t)\beta^T_\lambda(u) - \beta_\lambda(t)\alpha_\lambda^T(u)\bigg)
+i\beta_\lambda(t)\bigg(\gamma_\lambda(T)\beta^T_\lambda(u) -\alpha_\lambda^T(u)\bigg),\cr
G_{\bar \zeta \zeta}(t,u) &=-i\theta(t-u)\bigg(\bar\beta_\lambda(t)\alpha_\lambda^\dagger(u) - \bar\alpha_\lambda(t)\beta_\lambda^\dagger(u)\bigg)
+i\bar\alpha_\lambda(t)\bigg(\gamma_\lambda(T)\alpha_\lambda^\dagger(u) -\beta_\lambda^\dagger(u)\bigg),\cr
G_{\bar \zeta\bar \zeta}(t,u) &=-i\theta(t-u) \bigg(\bar\beta_\lambda(t)\beta_\lambda^T(u) - \bar\alpha_\lambda(t)\alpha_\lambda^T(u)\bigg)
+i\bar\alpha_\lambda(t)\bigg(\gamma_\lambda(T)\beta_\lambda^T(u) -\alpha_\lambda^T(u)\bigg).
\cr
}
\tag
$$
We then have
$$
\eqalign{
{\rm Tr}(G_\lambda N) &= - \int_0^T dt\, {\rm tr}\left\{ G_{\zeta\bar \zeta}(t,t)B(t)  +  G_{\bar \zeta \zeta}(t,t)\bar B(t)\right\}\cr
&=-i\,  \int_0^T dt\,{\rm tr}\Big\{\beta_\lambda(t)\bigg(\gamma_\lambda(T)\beta^T_\lambda(t) -\alpha_\lambda^T(t)\bigg)B(t)
+\bar\alpha_\lambda(t)\bigg(\gamma_\lambda(T)\alpha_\lambda^\dagger(t) -\beta_\lambda^\dagger(t)\bigg)\bar B(t)\Big\}.
}\tag trgn
$$

Equation \(trgn) can be simplified considerably. In particular, we  claim that 
$$
\Tr(G_\lambda N) = \tr(\bar\alpha_\lambda^{-1}(T){d\over d\lambda}\bar\alpha_\lambda(T)),
\tag tracegn
$$
where the trace on the right hand side of the equation is on $\V_-$. 
To prove \(tracegn), we vary the equations \(slambdaeq) determining $\Sym_\lambda$ so that
$$
\left({d\over dt} + H_\lambda\right){d\Sym_\lambda\over d\lambda} + {dH_\lambda\over d\lambda}\Sym_\lambda = 0.
\tag eqdsdlambda
$$ 
Equation \(eqdsdlambda) can be viewed as an inhomogeneous equation for $dS_\lambda/d\lambda$ with ``source'' given by $-{dH_\lambda\over d\lambda}\Sym_\lambda$. With initial condition
$$
\left({d\Sym_\lambda\over d\lambda}\right)_{t=0} = 0,
\tag
$$
the solution is
$$
{d\Sym_\lambda(t)\over d\lambda} = -\int_0^t du\, \Sym_\lambda(t) \Sym_\lambda^{-1}(u) {dH_\lambda(u)\over d\lambda}\Sym_\lambda(u).
\tag dsym
$$
Using \(symlambda) and
$$
{dH_\lambda\over d\lambda} = i\left(\matrix{
0&\bar B\cr
- B&0
}\right),
\tag
$$
in \(dsym) we get
$$
\eqalign{
{d\bar \alpha_\lambda(t)\over d\lambda} &= i\int _0^t du\,\Bigg\{\Big[ \bar\alpha_\lambda(t)\alpha_\lambda^T(u) - \bar\beta_\lambda(t)\beta_\lambda^T(u)\Big] B(u)\beta_\lambda(u)
\cr
&\phantom{ i\int _0^t du\,\Bigg\{\Big[ }
\quad+\Big[\bar\alpha_\lambda(t)\beta_\lambda^\dagger(u) - \bar\beta_\lambda(t)\alpha_\lambda^\dagger(u)\Big]\bar B(u)\bar \alpha_\lambda(u)\Bigg\}.}
\tag
$$
We then have
$$
\eqalign{
\tr\left(\bar\alpha_\lambda^{-1}(T) {d\bar \alpha_\lambda(T)\over d\lambda}\right)
&=  i\int _0^T du\,\tr\Bigg\{\Big[\alpha_\lambda^T(u) - \gamma_\lambda(T)\beta^T(u)\Big] B(u)\beta_\lambda(u)
\cr
&\phantom{=  i\int _0^T du\,\Bigg\{\Big[}
+\Big[\beta_\lambda^\dagger(u) - \gamma_\lambda(T)\alpha^\dagger(u)\Big]\bar B(u)\bar \alpha_\lambda(u)\Bigg\}
\cr
&=\Tr\left(D^{-1}_\lambda N\right).}
\tag
$$

With this result in hand we can obtain the final form for the path integral. Using
$$
\tr\left(\bar\alpha_\lambda^{-1}(T) {d\bar \alpha_\lambda(T)\over d\lambda}\right) = {d\over d\lambda}\big[ \log\det (\bar\alpha_\lambda(T))\big],
\tag
$$
and  \(nude) we have
$$
{d\over d\lambda} \Big\{\log {\I}_\lambda +\half \log\det(\bar\alpha_\lambda(T))\Big\} = 0,
\tag
$$
so that, using \(ibc), we have
$$
\det^{-\half}(D_0^{-1} D) = {\cal I}_1 = \det^{-\half}(\bar\alpha_0^{-1}(T)\bar\alpha_1(T)).
\tag
$$
Finally, from \(theintegral) and $\alpha_1=\alpha$ we have
$$
\I(v,w) =  \expiS
\sqrt{1\over \det(\bar\alpha_0^{-1}(T)\bar\alpha(T))}.
\tag finalans
$$
We note that the coherent state path integral result \(finalans) for the transition amplitude has been rigorously obtained within the Fock space formalism for a particular class of normal-ordered, time-independent, quadratic Hamiltonians \refto{Honegger1996}.

\taghead{5.}
\head{5. Discussion}

Evidently, the path integral as computed in \(finalans) makes sense provided the exponential factor and the determinant of the operator $K=\alpha_0^{-1}(T)\alpha(T)$ exist. For the Fredholm determinant of an operator $K$ to be defined it is necessary and sufficient that $K-id$ is a trace-class operator. (In this section ``$id\,$'' denotes the identity operator on $\V_+$.)  An important necessary condition for $\det(K)$ to exist is that the Bogoliubov coefficient $\beta(T)$ in $\Sym(T)$ defines a Hilbert-Schmidt operator:
$$
{\rm tr} [\beta^\dagger(T)\beta(T)]   <\infty.
\tag
$$  
This can be seen by first noting that $\alpha_0$ is a unitary operator, which follows from the easily established fact that $\beta_0=0$. Then we have 
$$
|\det(K)|^2 = \det(K K^\dagger) = \det(id + \alpha_0^{-1}\beta\beta^\dagger\alpha_0),
\tag
$$
where all operators are evaluated at time $T$.  The operator $\alpha_0\beta\beta^\dagger\alpha_0^\dagger$ is trace-class if and only if $\beta\beta^\dagger$ is, \ie $\beta(T)$ must be Hilbert-Schmidt. When $\beta(T)$ is Hilbert-Schmidt it follows that the exponential factor in \(finalans) exists because each of $\alpha^{-1}(T)$, $\sigma(T)$ and $\gamma(T)$ is bounded.
 This necessary condition is highlighted here because it is known that the symplectic transformation $\Sym(T)$ is unitarily implementable in the Fock space representation defined by $\Omega$ and $J$ if and only if $\beta(T)$ is Hilbert-Schmidt \refto{Shale1962, Honegger1996}.\footnote*{There are are number of interesting situations where this Hilbert-Schmidt condition fails (see, \eg \reftorange{CGT1999c}{Gowdy, Helfer1996} {Helfer1999}).} Thus the path integral, normalized using $Q_0$ and interpreted using the Fredholm determinant, fails to exist if the symplectic transformation generated by the classical Hamiltonian fails to be unitarily implementable in the Fock space representation. 
 
It should be emphasized, however, that even if the symplectic transformation corresponding to time evolution from $t=0$ to $t=T$ {\it is} unitarily implemented, this does not guarantee that the path integral exists. Implementability means  only that  the absolute value of the determinant is defined --- the phase of the determinant may not be defined.

In the case where the Hamiltonian is time-independent, some sufficient conditions for the Fredholm determinant of $K$ to exist (and hence for the exponential factor to exist as well) can be obtained from the results in \refto{Berezin1966, Honegger1996}. For example, a relatively simple sufficient condition is that the quadratic form on $\V_+$ given by $B$ corresponds (via the scalar product on $\V_+$) to a Hilbert-Schmidt operator:
$$
\sum_{ab} B_{ab}\bar B^{ab} <\infty.
\tag
$$
From the point of view of the Fock space formulation of the quantum system, this implies that the symplectic transformations $\Sym(t)$ form a strongly continuous group and are represented (projectively) as a continuous unitary group generated by a Hamiltonian operator, which is unique up to addition of a multiple of the identity. The Hilbert-Schmidt condition on $B$ is then equivalent to requiring that the vacuum state of the Fock representation defined by $\Omega$ and $J$ is in the domain of the Hamiltonian. 

\bigskip\noindent{\bf Acknowledgments}

This work was supported in part by National Science Foundation grant PHY-0244765 to Utah State University.

\refis{Helfer1996}{A. Helfer, \journal Class. Quantum Grav., 
13, L129, 1996.}

\refis{Helfer1999}{A. Helfer, preprint hep-th/9908011.}

 \refis{CGT1999c}{C. G. Torre and M. Varadarajan, \journal 
 Class. Quantum Grav., 16, 2651, 1999.}
 
\refis{Wald1994}{R. Wald, {\it Quantum Field Theory in 
Curved Spacetime and Black Hole Thermodynamics}, (University 
of Chicago Press, Chicago 1994).}

\refis{Shale1962}{D. Shale, {\it Trans. Am. Math. Soc.}, {\bf 103}, 149 (1962).}

\refis{Baranger2001}{M. Baranger, {\sl et al}, \journal {\sl J.~Phys.~A}, 34, 7227, 2001.}

\refis{Itzykson1980}{C.~Itzykson and J.~Zuber, {\it Quantum Field Theory}, (McGraw Hill, New York, 1980).}

\refis{Faddeev1980}{L.~Faddeev and A.~Slavnov, {\it Gauge Fields: Introduction to Quantum Theory} (Benjamin Cummings, 1980).}

\refis{Schulman1981}{L.~Schulman, {\it Techniques and Applications of Path Integration}, (Wiley, New York, 1981).}

\refis{Klauder1985}{J.~Klauder and B.~Skagerstam, {\it Coherent States}, (World Scientific, Singapore, 1985).}

\refis{Honegger1996}{R. Honegger and A. Rieckers, JMP 37, 4292 (1996).}

\refis{Bratteli1979}{O. Bratteli and D. Robinson, Operator Algebras and Quantum Statistical Mechanics (Springer, 1979).}

\refis{DeWitt1995}{P. Cartier and C. DeWitt-Morette, JMP 36, 2137 (1995).}

\refis{Albeverio1976}{A. Albeverio and R. Hoegh-Krohn, Mathematical Theory of Feynman Path Integrals, (Springer, 1976).}

\refis{Berezin1966}{F. Berezin, The Method of Second Quantization (Academic Press, New York, 1966).}

\refis{MV2004}{M.~Varadarajan, \prd  70, 084013, 2004.}

\refis{Gowdy}{A. Corichi, {\it et al}, \journal {\sl Int.~J.~Mod.~Phys. D}, 11, 1451, 2002; C.~G.~Torre, \prd 66, 084017, 2002.}

\references

\endreferences

\bye